\documentclass[final,5p,times,twocolumn]{elsarticle}

\usepackage{amssymb}
\usepackage{mathptmx}
\usepackage{amsthm}
\usepackage{graphicx}
\graphicspath{{Figures/}{NewFigures/}}
\usepackage{xcolor}
\usepackage{floatrow}
\usepackage{natbib}
\usepackage{balance}
\biboptions{sort&compress}
\usepackage{hyperref}
\hypersetup{
  colorlinks   = true, 
  urlcolor     = blue, 
  linkcolor    = blue, 
  citecolor   = blue 
}
\usepackage[pagewise]{lineno}

\journal{Combustion and Flame}

\begin{document}

\begin{frontmatter}



\title{Critical region in the spatiotemporal dynamics of a turbulent thermoacoustic system and smart passive control}

\author[label1]{Amitesh Roy\corref{cor1}}
\ead{amiteshroy94@yahoo.in}
\cortext[cor1]{Corresponding Author}
\author[label2]{C P Premchand}
\author[label1]{Manikandan Raghunathan}
\author[label1]{Abin Krishnan}
\author[label2]{Vineeth Nair}
\author[label1]{R I Sujith}
\address[label1]{Department of Aerospace Engineering, IIT Madras, Chennai, Tamil Nadu - 600 036, India}
\address[label2]{Department of Aerospace Engineering, IIT Bombay, Mumbai, Maharashtra - 400 076, India}

\begin{abstract}
We develop a passive control strategy for suppressing thermoacoustic instability in a bluff-body stabilized premixed turbulent combustor. When the equivalence ratio is varied, there is a transition from combustion noise to thermoacoustic instability via intermittency in the combustor. We perform simultaneous acoustic pressure, 2D-PIV, and CH$^*$ chemiluminescence measurements to capture the pressure fluctuations, the velocity field, and the heat release rate (HRR) field during the transition. We measure the spatial distribution of the amplitude of turbulent velocity at the acoustic frequency, time-averaged vorticity, time-averaged HRR, and Rayleigh index and identify various regions of significance. We implement a passive control strategy by targeting these regions with a steady injection of secondary micro-jet of air to optimize the injection location and determine the critical region. Targeting the critical region with secondary air leads to greater than 20 dB suppression of the dominant thermoacoustic mode. We observe that the coherent structure forming from the shear layer following the dump plane gets suppressed, leading to an incoherent spatial distribution of HRR fluctuations. We find that the turbulent velocity amplitude correctly identifies the critical region for optimized passive control during thermoacoustic instability. In contrast, the Rayleigh index identifies the region of the most significant acoustic driving; however, it does not identify the region most sensitive to control. Finally, we extend our analysis by determining the spatial distribution of the Hurst exponent measured from the turbulent velocity field. We show that the Hurst exponent identifies the critical region during thermoacoustic instability and intermittency, unlike the other physical measures. Thus, we develop a smart passive control method by combining the need for finding critical regions in the combustor with the predictive capabilities of the Hurst exponent.
\end{abstract}


\end{frontmatter}


\section{Introduction}
\label{Introduction}

Thermoacoustic instability refers to large-amplitude periodic pressure oscillations arising from the positive feedback between unsteady combustion and the acoustic modes of the combustion chamber \cite{lieuwen2005combustion}. The problem is exacerbated in land-based gas turbine combustors, which {\color{black}operate} in fuel-lean conditions to reduce NOx emission and meet environmental norms. Such a susceptibility of lean combustion systems to thermoacoustic oscillations results from the high sensitivity of lean premixed flames to harmonic perturbations arising from pressure oscillations and flow instabilities \cite{lieuwen2003modeling}.  Each year, thermoacoustic instability causes billions of dollars of revenue loss directly through repair and replacement costs of failed combustors in gas turbine industries and indirectly due to combustor downtime and associated power outages in power plants \cite{lieuwen2005combustion}. The complex interactions between turbulence, combustion, and acoustics of the combustion chamber lead to a variety of nonlinear behavior \cite{juniper2018sensitivity} and have necessitated the use of different approaches in making the problem tractable. These include the use of flame transfer and describing functions \cite{schuller2020dynamics}, distributed time-lag models \cite{polifke2020modeling}, adjoint methods \cite{magri2019adjoint} and complex systems approach \cite{sujith2020complex} in characterizing and controlling thermoacoustic oscillations.

\subsection{Transition to thermoacoustic instability}
During stable combustor operation, the sound generated from turbulent flames is due to non-steady volumetric expansion and convective entropy modes \cite{candel2009flame}. The radiated sound lacks any characteristic time scale and has a broadband signature, and is referred to as combustion noise \cite{candel2009flame}. Later studies revealed that combustion noise displays scale invariance \cite{murugesan2015combustion} and possess signatures of multifractality \cite{nair2014multifractality}. During unstable combustor operation or thermoacoustic instability, feedback between the heat release rate (HRR) and the acoustic pressure fluctuations lead to large-amplitude periodic pressure oscillations. By systematically varying the control parameters, it is possible to transition from stable to unstable combustor operation. In turbulent combustors, the transition is often associated with an intermediate state known as intermittency  \cite{nair2014intermittency}. Intermittency is characterized by the presence of periodic bursts interspersed with chaotic oscillations. During the transition to thermoacoustic instability, there is a gradual loss in multifractality associated with the acoustic pressure and HRR oscillations \cite{nair2014multifractality, unni2015multifractal}. In other words, there is a transition from a state possessing multiple time scales to one possessing a single characteristic time scale. 

The flow field also undergoes drastic changes during the transition in a bluff-body stabilized combustor. {\color{black}During the occurrence of combustion noise and aperiodic epochs of intermittency, small vortices are shed aperiodically, and the HRR field remains spatially incoherent. In contrast, during thermoacoustic instability and periodic epochs of intermittency, vortices are shed periodically from the combustor's backward-facing step. These vortices carrying the air and fuel mixtures develop into large-scale coherent structures, which upon impingement with the bluff-body and combustor walls, result in regions with intense HRR \cite{george2018pattern, poinsot1987vortex}. Raghunathan et al. \cite{raghunathan2020multifractal} measured the multifractal spectrum from the spatial distribution of wrinkles on the flame surface at different states during the transition. They showed that the span of the multifractal spectra increases during thermoacoustic instability, indicating the significant increase in the spatial scales over which HRR fluctuations occur \cite{raghunathan2020multifractal}.}

Thus, the dynamics of a thermoacoustic system are controlled by the spatiotemporal evolution of the interaction between three subsystems -- turbulent flow field, combustion, and acoustic field of the combustor. Quantifying this spatiotemporal evolution using the right mathematical tool is crucial in forewarning an impending thermoacoustic instability. Nair et al. \cite{nair2014intermittency} showed that the scaling of the acoustic pressure oscillations (quantified by the Hurst exponent, $H$) during combustion noise and intermittency fares better than corresponding measures based on tracking the amplitude of the pressure oscillations when predicting thermoacoustic instability. However, such single point acoustic measurements cannot be used to predict in advance the changes required in the combustor to control thermoacoustic instability. Thus, there is a need for extending such an analysis in the spatial domain. 

\subsection{Control of thermoacoustic instability}
The different approaches used for controlling thermoacoustic instability can be broadly classified into passive and active control strategies. In passive control, some aspect of the combustor (geometry, injector arrangement, {\color{black}dampers} etc.) is changed independent of the operation of the {\color{black}combustor \cite{zhao2015review}.} Passive control strategies are widely implemented in practical combustion systems {\color{black} as they require low maintenance and are highly durable.}

Active control involves continuously monitoring the combustor and taking control measures based on the specific state of the system \cite{zhao2018review}. These involve actuators and tunable valves modulating the primary and secondary air and fuel flow rates and has been successfully used for suppressing thermoacoustic instability \cite{langhorne1990practical, mcmanus1990combustor, uhm2005low, hathout2002combustion}. Modulation of secondary fuel or air relies crucially on the response of high-speed actuators being robust. Ensuring fast response of actuators at frequencies where the combustor dynamics are most sensitive to forcing is a challenge and is the main drawback in the implementation of active control strategies. 

Motivated by these limitations, Ghoniem and co-workers \cite{ghoniem2005stability, altay2007impact, altay2010mitigation} considered steady injection of secondary air for achieving control in a dump combustor. Ghoniem et al. \cite{ghoniem2005stability} found that the momentum-ratio of the jet to the main flow above unity leads to a compact flame structure that is less driven by the wake vortex. Later, Altay et al. \cite{altay2010mitigation} compared the suppression observed during transverse and streamwise secondary air injection. For optimum transverse injection, a compact flame structure anchored upstream of the backward-facing step led to the suppression. In contrast, optimum streamwise injection inhibited unsteady vortex formation at the backward-facing step, leading to suppression.

Next, we consider the importance of the location of secondary air injection. Although injecting secondary air near the location of flame anchoring leads to suppression, it also affects the flame stability. Consequently, Ghoniem et al. \cite{ghoniem2005stability} and Altay et al. \cite{altay2010mitigation} {\color{black}used secondary $H_2$ injection} to increase flame stability and prevent blow-out. However, injecting $H_2$ led to higher flame temperatures and increased NOx levels considerably. Thus, Ghoniem et al. \cite{ghoniem2005stability} had to optimize the main flow after the injection of secondary air and $H_2$ to reduce the temperature inside the combustor and decrease the NOx levels. Later, Oztarlik et al. \cite{oztarlik2020suppression} showed that secondary $H_2$ injection in small fractions alone could suppress TAI. 

To by-pass the back and forth adjustments in the main and secondary airflow, prevent flame blow-out, and reduce the complexity involved in maintaining expensive $H_2$ plumbing, we consider steady and unmodulated injection of secondary air away from the region of flame anchoring. The optimal region for injection can be selected if we can identify local regions responsible for TAI. Uhm and Acharya \cite{uhm2005low} {\color{black}considered} the region of local maxima in HRR to be the optimum region. In contrast, Ghoniem and co-workers \cite{ghoniem2005stability, altay2007impact, altay2010mitigation} rationalized the optimum region as the region of flame anchoring, which led to flame stabilization problems. In a similar study, Tachibana et al. \cite{tachibana2007active} used a distribution of the Rayleigh index to optimize for the choice of the secondary fuel injector. Recently, Unni et al. \cite{unni2018emergence} used network centrality measures derived from the Pearson correlation coefficient to identify regions of critical importance during thermoacoustic instability for the bluff-body stabilized combustor used in this study. In a follow-up study, Krishnan et al. \cite{abin2019critical} demonstrated that targeting regions with large network measure values leads to the most effective control of thermoacoustic instability. 

\subsection{Scope}
In the discussion above, we highlighted the importance of quantifying the spatiotemporal behavior of the reacting flow field during the transition to the state of TAI. We also emphasized the challenges in developing even a relatively simple passive control strategy involving an unmodulated secondary micro-jet of air. Thus, the key objectives of this study are:
\begin{enumerate}
\item To optimize for secondary air-injection location targeting the \textit{critical region} in the combustor without hampering the flame stability. Such a critical region is determined from spatiotemporal quantities such as the amplitude of velocity fluctuations, mean vorticity, Rayleigh index, and averaged HRR field.

\item To extend the analysis using Hurst exponent ($H$) in the spatial domain. As mentioned earlier, $H$ measured from single-point pressure measurements is invaluable in predicting impending instability. A spatial analysis using $H$ may provide useful insights in optimizing secondary air injection location during the states leading up to TAI. Consequently, the advantages of optimized passive control can be complemented by the predictive capability of the Hurst exponent resulting in a smart passive control strategy. 
\end{enumerate}

The manuscript is structured as follows: The experimental setup with the turbulent combustor is described in \S\ref{2.1. Experimental setup and measurements} and the combustor is characterized in \S\ref{2.2. Characterizing the turbulent combustor}. The methodology of determining the spatial distribution of $H$ from the flow field is described in \S\ref{2.3. Tools for analysis}. In \S\ref{3.1. Spatio-temporal}, we discuss the spatiotemporal dynamics through measures based on the amplitude of velocity fluctuations, vorticity, averaged HRR, and Rayleigh index. In \S\ref{3.3. Smart passive control}, we discuss passive control based on the information from the physical measures discussed in \S\ref{3.1. Spatio-temporal}. In \S\ref{3.4 Spatial Hurst exponent}, we discuss the results based on the distribution of $H$ and highlight the key advantages of using $H$ over physical measures. Lastly, in \S\ref{4. Conclusions}, we emphasize the key contributions of this study in conclusion.

\section{Methodology}
\label{2.Methodology}

\subsection{Experimental setup and measurements}
\label{2.1. Experimental setup and measurements}

\begin{figure}[t!]
\centering
\includegraphics[width=\textwidth]{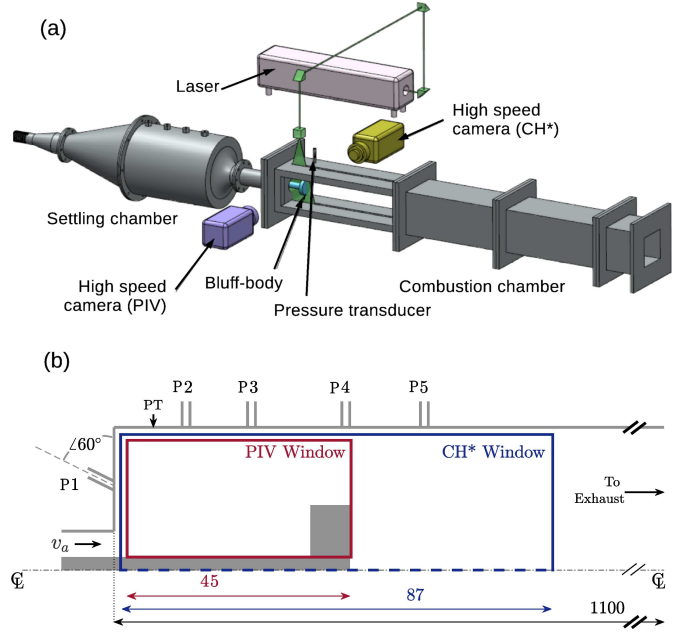}
\caption{\label{Fig1_expt_setup} (a) The bluff-body stabilized turbulent combustor used for the present study. (b) Schematic of the combustor cross-section indicating the PIV and CH$^*$ field of view. Secondary air injection ports (P1-P5) are present on either side of the centerline for the passive control study. PT indicates the location of the pressure transducer. All dimensions are in mm. Reproduced and adapted with permission from \cite{krishnan2019emergence}.}
\end{figure}

We perform experiments in a bluff-body stabilized turbulent combustor {\color{black}at atmospheric conditions} (Fig. \ref{Fig1_expt_setup}a). The turbulent combustor has a cross-section of $90\times90$ mm$^2$ and is 1100 mm long. The bluff-body comprises a circular disk of 47 mm diameter and 10 mm thickness. It is centrally mounted on a hollow shaft of $d_s=16$ mm diameter. The bluff-body is located 35 mm from the dump plane of the combustor (Fig. \ref{Fig1_expt_setup}b). Air first passes through a settling chamber before being guided into the combustor through an inlet of diameter $d_i=40$ mm. Fuel (Liquified Petroleum Gas or LPG, $60\%$ butane and $40\%$ propane) is injected through holes of 1.7 mm diameter present on the hollow shaft, 110 mm upstream of the backward-facing step. {\color{black}The expansion ratio of the combustor is $6.45$. The partially premixed fuel-air mixture is ignited using a spark plug connected to an 11 kV transformer and mounted on the dump plane. The combustion products are exhausted through a decoupling chamber (1000 mm $\times$ 500 mm $\times$ 500 mm) into the atmosphere.}

Air and fuel flow rates are controlled through mass flow controllers (Alicat Scientific, MCR series) and have a measurement uncertainty of $\pm$($0.8\%$ of reading + $0.2\%$ of full-scale). In our experiments, the fuel flow rate was maintained at 0.95 g/s, and the air flow rate was varied from 9.80 g/s {\color{black}to} 15.92 g/s such that the equivalence ratio ($\phi$) varied in the range of 0.95 to 0.53. The air velocity varies in the range of $\upsilon_a=8.1$ m/s to 14 m/s. The air flow Reynolds number, calculated as $Re=\upsilon_a(d_i-d_s)/\nu$, varies from 12500 to 18000, where $\nu$ is the kinematic viscosity of air. The maximum uncertainty in the indicated value of $\phi$, $\upsilon_a$, and $Re$ are $\pm 1.6\%$, $\pm 0.8\%$ and $\pm0.8\%$, respectively.

Secondary air injection ports of 5 mm diameter are present on either side of the centerline, as shown in Fig. \ref{Fig1_expt_setup}b. {\color{black}The injection port P1 is mounted on the dump plane and is located at a distance of $23.5$ mm from the centerline. The four transversely mounted injection ports (P2-P5) are located at 15 mm, 25 mm, 45 mm, and 65 mm, respectively, from the dump plane.} These ports target different regions of the flow field and are used for passive control. The injection ports are mounted differently from previous studies \cite{ghoniem2005stability, altay2010mitigation, uhm2005low}, which targeted the flame anchoring point. This particular secondary injection configuration was chosen to avoid flame anchoring and stability problems during control experiments. The secondary air is controlled through a separate mass flow controller. Secondary air is injected into the combustor during the state of thermoacoustic instability in steps of 0.16 g/s {\color{black}until} 1.90 g/s.

Pressure measurements are performed using a PCB103B02 piezoelectric transducer (sensitivity: 217 mV/kPa and uncertainty: $\pm0.15$ Pa) mounted on the combustor wall, 17 mm from the dump plane. Two-component 2-D high-speed particle image velocimetry (PIV) is performed to acquire the velocity field. The reactive flow field is seeded using 1 $\mu$m diameter $TiO_2$ particles. Mie scattering images were acquired using a high-speed CMOS camera (Photron SA-4). The procedure for determining the velocity field from Mie scattering images is detailed in \cite{krishnan2019emergence}. The camera is equipped with a ZEISS 100 mm camera lens at $f/2$ aperture. Chemiluminescence images are captured using Phantom - V12.1 with a ZEISS 50 mm camera lens and outfitted with a bandpass filter centered around $435\pm10$ nm to capture the emissions from CH$^*$ radicals from a region spanning $87\times78$ mm around the bluff-body (Fig. \ref{Fig1_expt_setup}a). The flow-field spanning $45\times 40$ mm between the dump plane and the bluff-body is imaged (Fig. \ref{Fig1_expt_setup}b). The sampling frequencies for pressure, chemiluminescence, and PIV measurements are 20 kHz, 4 kHz, and 2 kHz, respectively.

We performed optical diagnostics at specific fuel and airflow rates, which corresponded to states representative of combustion noise, intermittency, and thermoacoustic instability. We also performed diagnostics during the control experiment to evaluate the effect of secondary jet on the flow and HRR characteristics. For a more detailed discussion on the experimental setup and uncertainty measurements associated with measurement devices, please refer to \cite{george2018pattern, abin2019critical}.

\subsection{Characterizing the turbulent combustor}
\label{2.2. Characterizing the turbulent combustor}
\begin{figure}[t!]
\centering
\includegraphics[width=\textwidth]{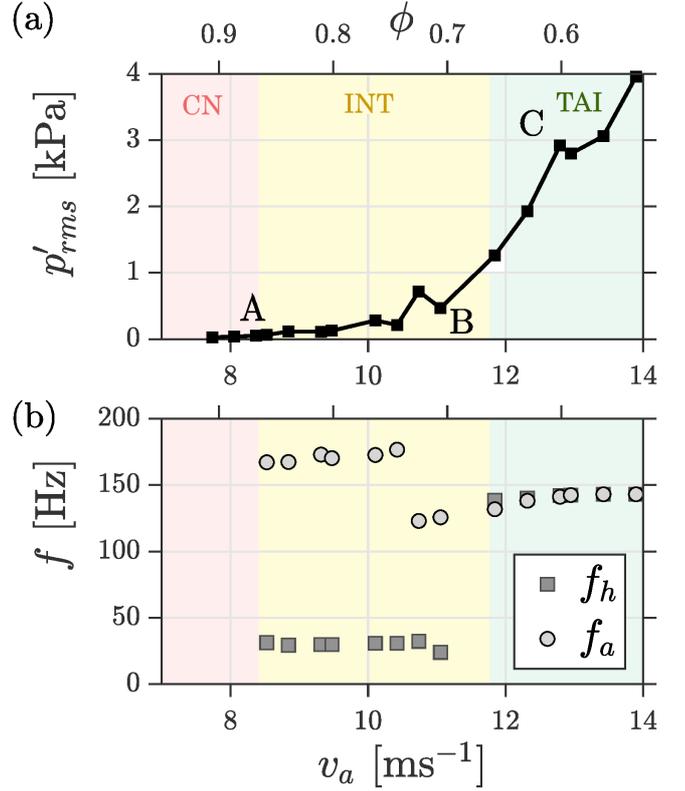}
\caption{\label{Fig2_prms_vs_ma} Intermittency route to thermoacoustic instability observed in the turbulent combustor. Variation of (a) $p^\prime_{rms}$ and (b) the frequency corresponding to the acoustic ($f_a$) and hydrodynamic mode ($f_h$) as a function of nominal velocity of air $\upsilon_a$ (bottom axis) and equivalence ratio $\phi$ (top axis). (a) has been adapted from \cite{george2018pattern} and (b) from \cite{premchand2019lagrangian_a}.}
\end{figure}

We depict the transition of the turbulent combustor from the state of combustion noise to thermoacoustic instability through the state of intermittency in Fig. \ref{Fig2_prms_vs_ma}. In Fig. \ref{Fig2_prms_vs_ma}a, we show the change in $p^\prime_{rms}$ as a function of the nominal velocity $\upsilon_a$ (bottom axis) of air and equivalence ratio $\phi$ (top axis). Markers `A', `B' and `C' correspond to three points representative of the states of combustion noise, intermittency, and thermoacoustic instability for the subsequent spatiotemporal analysis. Figure \ref{Fig2_prms_vs_ma}b shows the change in the dominant frequency of the acoustic pressure and spatially-averaged velocity time series during the transition to thermoacoustic instability. {\color{black} We can observe the two separate time scales -- acoustic ($1/f_a$) and hydrodynamic ($1/f_h$) -- during intermittency \cite{premchand2019lagrangian_a}. At the onset of thermoacoustic instability, the frequency of vortex shedding matches with the acoustic mode of the combustor. This process is associated with the mutual synchronization of the acoustic, hydrodynamic, and HRR field of the combustor \cite{pawar2017thermoacoustic, premchand2019lagrangian_a}.} In \S\ref{3.1. Spatio-temporal}, we discuss the spatiotemporal dynamics from the perspective of optimization for the location of secondary injection. A more thorough discussion on the spatiotemporal dynamics of bluff-body stabilized combustors can be found elsewhere \cite{poinsot1987vortex, chakravarthy2007vortex, george2018pattern, premchand2019lagrangian_a, premchand2019lagrangian_b}.

\begin{figure*}[t!]
\centering
\includegraphics[width=0.85\textwidth]{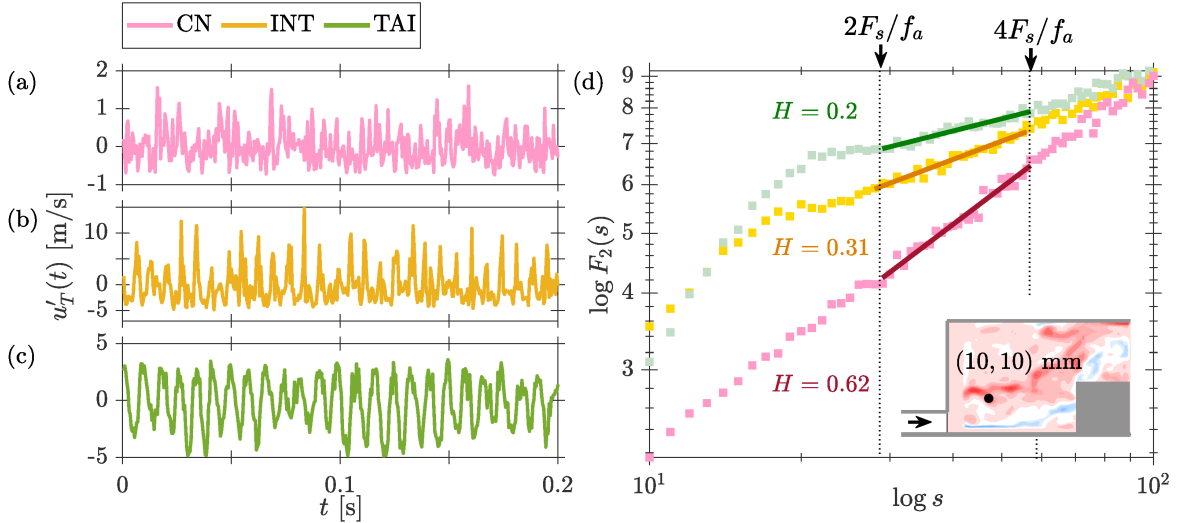}
\caption{\label{Fig3_Structure_Function_Scaling} (a-c) Time series of  turbulent velocity fluctuations ($u^\prime_T$) during combustion noise, intermittency and thermoacoustic instability at the indicated point inside the combustor shown in the inset. The three states correspond to point A, B and C indicated in Fig. \ref{Fig2_prms_vs_ma}. (d) The variation of the second-order structure function $F_2(s)$ measured from $u^\prime_T$ with the scale $s$ for the three states.}
\end{figure*}

\subsection{Nonlinear time series analysis: {\color{black}the} Hurst exponent}
\label{2.3. Tools for analysis}
The time-series of velocity fluctuations obtained from different locations in the combustor are generally non-stationary, and hence, central moments diverge over time \cite{nair2014intermittency}. Consequently, it is instructive to calculate the scaling of the moments on the time interval. The scaling exponent is called the Hurst exponent. The scaling is a measure of long-term memory in the system. Historically, the Hurst exponent was utilized to determine the optimum dam size for the Nile river based on long term flood and drought data \cite{hurst1951long}. The Hurst exponent was later connected to the geometry of fractals by Mandelbrot \cite{mandelbrot1982fractal}. We measure the Hurst exponent using Multifractal Detrended Fluctuation Analysis (MFDFA) \cite{kantelhardt2002Multifractal}, which we briefly discuss below.

{\color{black}We first perform Reynolds decomposition and calculate the fluctuations in $x$ and $y$ velocity component ($u_x^\prime$, $u_y^\prime$) by subtracting the respective time-averaged values ($\bar{u}_x$, $\bar{u}_y$) at all points in the velocity field obtained from PIV. We then determine the total magnitude of velocity fluctuations as:
\begin{equation}
u_T(x,y,t)= \sqrt{[(u_x^\prime(x,y,t)])^2+[u_y^\prime(x,y,t)]^2}.
\end{equation}}
Finally, we detrend the time series by subtracting the mean to obtain the fluctuations of the resultant velocity: $u^\prime_T=u_T-\bar{u}_T$, where $\bar{u}_T$ is the time-averaged resultant velocity. Note that $u^\prime_T$ is essentially the same as turbulent velocity fluctuations defined as the root mean square of the fluctuating velocity components. We plot the representative turbulent velocity fluctuations, measured at the indicated point in the inset, during the three states of combustor operation in Figs. \ref{Fig3_Structure_Function_Scaling}a-c. We can observe an increase in periodicity of the velocity fluctuations during the transition to thermoacoustic instability. 

We then calculate the cumulative deviate series: 
\begin{equation}
y_k(t) = \sum_{t=1}^k u_T^\prime(t)=\sum_{t=1}^k[u_T(t)-\bar{u}_T].
\end{equation}
The deviate series is divided into $n_s$ non-overlapping segments ($y_i(t)$, $i=1,2 ...n_s$) of equal span $s$. Local trends are removed by a local polynomial fit $\bar{y}_i$ onto the deviate series $y_i$. Local fluctuations are obtained by subtracting the fit from the deviate series. {\color{black}The computation time of the Hurst exponent is approximately proportional to the polynomial order used for local detrending. Hence, we consider a local linear fit for detrending as it is computationally faster and produces reliable results.} The $q^{th}$ order structure-function, $F_q$ can be obtained from the local fluctuations as \cite{kantelhardt2002Multifractal}:
\begin{equation}
F_q(s) = \Bigg[\frac{1}{n_s} \sum_{i=1}^{n_s}\Bigg(\frac{1}{s}\sum_{t=1}^{s}[y_i(t)-\bar{y}_i]^2\Bigg)^{q/2}\Bigg]^{1/q} \forall q\neq 0.
\end{equation}
The second-order structure function scales as $F_2(s)\sim s^H$ within the bounds set by the minimum and maximum window size. The Hurst exponent $H$ is then determined as:
\begin{equation}
H = \frac{\log F_2(s)}{\log s} \enspace \forall \enspace s\in[2/f_a,4/f_a].
\end{equation}

The Hurst exponent, $H$, measures the correlation and persistence in a time series. If a large (small) value is more likely to be followed by another large (small) value, the signal is said to be persistent and long-range correlated. Such signals have $H>0.5$. If a large (small) value is more likely to be followed by a small (large) value, the signal is anti-persistent. For such signals, $H<0.5$ and only short-range correlations exist \cite{nair2014multifractality}. $H=0$ indicates a purely periodic signal, while $H=0.5$ indicates white noise. Thus, the spatial distribution of $H$ indicates the persistence and correlation in the velocity fluctuations over the flow-field.

We plot the second-order structure function $F_2$ as a function of the scale $s$ measured from the time series of turbulent velocity fluctuations $u^\prime_T(t)$ at a representative point in the flow field for the three states of combustor operation in Fig. \ref{Fig3_Structure_Function_Scaling}d. We observe that the Hurst exponent is $H=0.62$ during combustion noise. The signal shown in Fig.\ref{Fig3_Structure_Function_Scaling}a is thus persistent and possesses long-range correlations. In fact, the signal is fractal in nature. In contrast, during intermittency and thermoacoustic instability, $H=0.31$ and $H=0.2$, which are much lower. The representative signal shown in Figs. \ref{Fig3_Structure_Function_Scaling}b,c suggests anti-persistent and periodic behavior. The decreasing value of $H$ during the transition to thermoacoustic instability is an indicator of the increasing periodic content in the system dynamics. We repeat this process over the entire flow field and obtain the spatial distribution of $H$ over the entire flow field.

We chose the scaling of the structure-function within the bounds $s\in[2/f_a,4/f_a]$ for two key reasons. First, $s\in[2/f_a,4/f_a]$ ensures that the range of the x-axis is not too low to be within a periodic cycle and not too large as to become completely uncorrelated \cite{kerres2016analysis}. Such behavior is also evident from Fig. \ref{Fig3_Structure_Function_Scaling}, where we observe that $F_2$ is relatively flat for low values of $s$ ($<2/f_a$), indicating the high correlation between the segments by them being in the same periodic cycle. Similarly, for large $s$ ($>4/f_a$), $F_2$ shows oscillations, indicating contributions from decorrelated segments. Second, the acoustics of the combustor {\color{black}impose} a characteristic length and time scale on the turbulent flow. Consequently, one only needs to be concerned with the scaling of the structure-function of turbulent velocity fluctuations in the range of time scales comparable to the acoustic time scale. The onset of thermoacoustic instability is captured by the gradual disappearance of the scaling of the structure-function in the bounded range of $s$.

\begin{figure*}[t!]
\centering
\includegraphics[width=\textwidth]{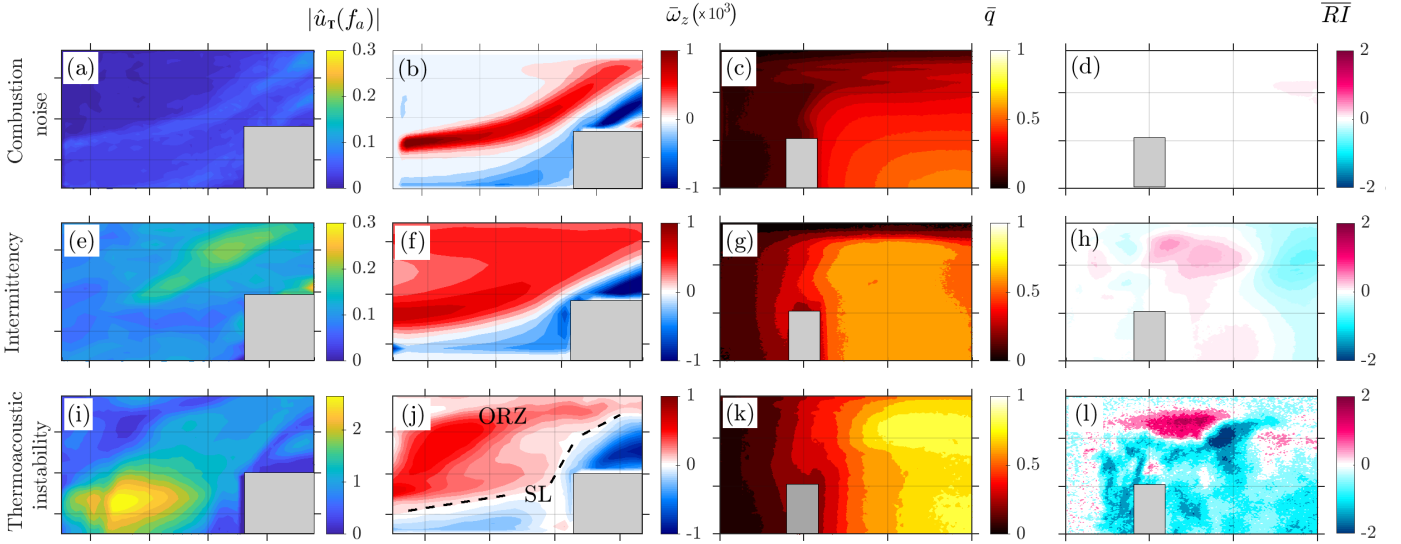}
\caption{\label{Fig4_CN_INT_TAI} Comparison of the Fourier amplitude of turbulent velocity fluctuations {\color{black}$|\hat{u}_T(f_a)|$} (m/s), time-averaged vorticity $\bar{\omega}_z$ ($\times 10^3$, s$^{-1}$), time-averaged heat release field $\bar{q}$ (a.u.) and time-average Rayleigh index $\overline{RI}$ (a.u.) during the states of (a-d) combustion noise, (e-h) intermittency and (i-l) thermoacoustic instability. The flow conditions  are indicated in Fig. \ref{Fig2_prms_vs_ma}. The span of the ordinate and abcissa are indicated in Fig. \ref{Fig1_expt_setup}b. The gray region indicates the position of the bluff-body.}
\end{figure*}

\section{Results and discussions}
\label{3. Results and discussions}

\subsection{Spatiotemporal analysis during the transition to thermoacoustic instability}
\label{3.1. Spatio-temporal}
We start by analyzing the spatiotemporal dynamics during the transition from combustion noise to thermoacoustic instability. As already discussed, we acquire simultaneous data associated with velocity, CH$^*$ chemiluminescence, and pressure fluctuations. In Fig. \ref{Fig4_CN_INT_TAI}, we plot the {\color{black}Fourier amplitude of the resultant velocity fluctuations ($u_T$) at the acoustic frequency  $f_a$ i.e., $|\hat{u}_T(f_a)|$,} the time-averaged vorticity ($\bar{\omega}_z(x,y)$), and the time-averaged heat release rate field ($\bar{q}(x,y)$) during combustion noise, intermittency and thermoacoustic instability at parametric points A, B and C indicated in Fig. \ref{Fig2_prms_vs_ma}a. We also plot the Rayleigh index, which is defined as:
\begin{equation}
\overline{RI}(x,y)=\frac{1}{NT}\int_0^{NT} p^\prime(t)\dot{q}^\prime(x,y,t)dt.
\end{equation}
Here, $N(=686)$ denotes the total number of cycles, and $T(=1/f_a)$ denotes the time-period of oscillations. The spatial distribution of the Rayleigh index quantifies the strength of acoustic power sources and sinks depending upon positive or negative feedback between pressure and heat release rate oscillations, respectively.

During the state of combustion noise at $\phi=0.86$ and $\upsilon_a=8.1$ m/s, flow fluctuations show broadband characteristics. The Fourier transformed amplitude {\color{black}$|\hat{u}_T(f)|$} at the acoustic mode $f=f_a$ is very low  (Fig. \ref{Fig4_CN_INT_TAI}a). Notice that the scale only extends {\color{black}until} 0.1 m/s, which is very low. The time-averaged vorticity field shown in Fig. \ref{Fig4_CN_INT_TAI}b indicates that the vortices evolve only along the shear layer (SL). The transverse span of vorticity contour indicates that the size of vortices is very small. We also notice the absence of the outer recirculation zone (ORZ). The time-averaged heat release rate field also shows very low values (Fig. \ref{Fig4_CN_INT_TAI}c). Likewise, the Rayleigh index shows very low values throughout the combustor (Fig. \ref{Fig4_CN_INT_TAI}d).

During the state of intermittency  at $\phi=0.66$ and $\upsilon_a=11.1$ m/s, {\color{black}$|\hat{u}_T(f_a)|$} shows higher values above the shear layer as compared to the state of combustion noise due to intermittent periodic oscillations of turbulent velocity induced by intermittent acoustic pressure oscillations (fig. \ref{Fig4_CN_INT_TAI}e). During epochs of periodic oscillations, larger vortices are shed, leading to the flow recirculating at the dump plane. This can be seen from the rather large distribution of $\bar{\omega}_z$ in Fig. \ref{Fig4_CN_INT_TAI}f. The maximum of $\bar{\omega}_z$ along the shear layer indicates that most of the vortices are shed along the shear layer but only recirculate intermittently. The time-averaged heat release rate field (Fig. \ref{Fig4_CN_INT_TAI}g) and Rayleigh index (Fig. \ref{Fig4_CN_INT_TAI}b) also show higher values when compared with that during combustion noise.

There is a significant change in the spatiotemporal behaviour during the state of thermoacoustic instability at $\phi=0.63$ and $\upsilon_a=12.3$ m/s (Figs. \ref{Fig4_CN_INT_TAI}i-l). Thermoacoustic instability in bluff-body stabilized combustors is associated with the phenomenon of vortex-acoustic lock-on wherein the frequency of vortex shedding matches the acoustic frequency $f_h=f_a$ (as shown in Fig. \ref{Fig2_prms_vs_ma}b) and is central to the establishment of the thermoacoustic feedback in the combustor \cite{poinsot1987vortex, chakravarthy2007vortex, pawar2017thermoacoustic, premchand2019lagrangian_a}. Accordingly, we observe a clearly defined region with a very high value of {\color{black}$|\hat{u}_T(f_a)|$} (Fig. \ref{Fig4_CN_INT_TAI}i). The periodically shed vortices from the dump plane and the tip of the bluff-body develop into large coherent structures which {\color{black}recirculate} into the outer recirculation zone (ORZ) and can be observed from the high value of $\bar{\omega}_z$ (Fig. \ref{Fig4_CN_INT_TAI}j).

\begin{figure*}[t]
\centering
\includegraphics[width=\textwidth]{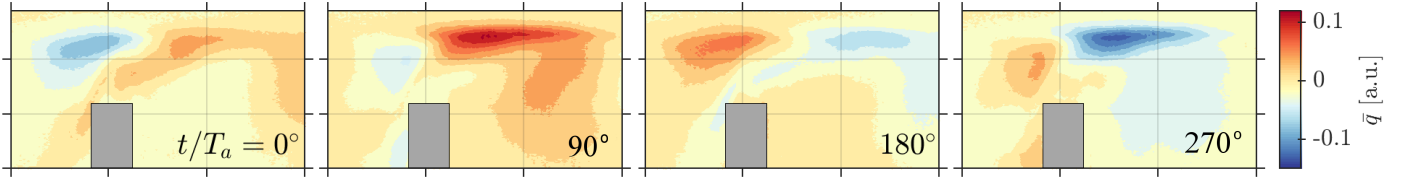}
\caption{\label{Fig5_Q_Ph_Avg.eps} Phase-averaged heat release rate field at the indicated phases of the acoustic cycle during the state of thermoacoustic instability. The gray mask indicates the position of the bluff-body.}
\end{figure*}

The large coherent structures carrying unburnt reactants impinge on the bluff-body and the combustor walls and lead to intense heat release. The time-averaged heat release rate field in Fig. \ref{Fig4_CN_INT_TAI}k shows maxima in the HRR fluctuations downstream of the bluff-body. To better understand the HRR dynamics, we show the phase-averaged value of HRR from the mean-subtracted chemiluminescence images at $0^\circ, 90^\circ, 180^\circ$, and $270^\circ$ of the acoustic cycle in Fig. \ref{Fig5_Q_Ph_Avg.eps}. The phase-averaged HRR field is indicative of the evolution of the flame structure at different points of the acoustic cycle. {\color{black} In the phase-averaged CH$^*$ field, we observe very intense fluctuations in the heat release rate above the bluff-body. The fluctuations intensify from moderate to high positive/negative values as one moves closer to maxima ($90^\circ$) and minima ($270^\circ$) of the pressure fluctuations. At pressure maxima ($90^\circ$), we observe large positive heat release rate fluctuations occupying considerable space above the bluff-body. During pressure minima ($270^\circ$), negative fluctuations dominate the field. This behavior is consistent with the Rayleigh criteria, where large positive fluctuations in the HRR field appear at pressure maxima and vice versa. The relation between pressure and HRR fluctuations is better understood from the plot of the local Rayleigh index in Fig. \ref{Fig4_CN_INT_TAI}l. We observe very high and positive values of the local Rayleigh index in the region above the bluff-body. These are the primary acoustic power sources responsible for thermoacoustic instability in the system.}

As we mention earlier, precursor based methods are great for predicting impending thermoacoustic instability. However, such methods do not provide any information required for implementing passive control measures. Passive control requires knowledge about the relative importance of different regions in the flow field. Based on the considerations made above, we find different regions of interest from different measures. For instance, the region between the dump plane and the bluff-body appears significant based on the large amplitude of {\color{black}$|\hat{u}_T(f_a)|$} during thermoacoustic instability (Fig. \ref{Fig4_CN_INT_TAI}i). Similarly, the maxima in $\bar{\omega}_z$ during thermoacoustic instability (Fig. \ref{Fig4_CN_INT_TAI}j) emphasizes the importance of the outer recirculation zone. Likewise, $\bar{q}(x,y)$ points towards the region after the bluff-body (Fig. \ref{Fig4_CN_INT_TAI}k), while phase-averaged heat release field (Fig. \ref{Fig5_Q_Ph_Avg.eps}) and Rayleigh index (Fig. \ref{Fig4_CN_INT_TAI}l) highlights the importance of the region above the bluff-body. We test the efficacy of passive control when these ``critical'' regions are selectively targeted with secondary air injection. Finally, as with the precursor, we would like to determine whether it is possible to predict such critical regions inside the combustor during the state of intermittency itself. 

\subsection{Passive control of thermoacoustic instability}
\label{3.3. Smart passive control}

We now attempt passive control of thermoacoustic instability using secondary air injection from various ports mounted on the combustor side-walls (see Fig. \ref{Fig1_expt_setup}b). Secondary air injected through these ports target various regions identified through the physical measures considered in the previous section. 

\begin{figure}[t!]
\centering
\includegraphics[width=\textwidth]{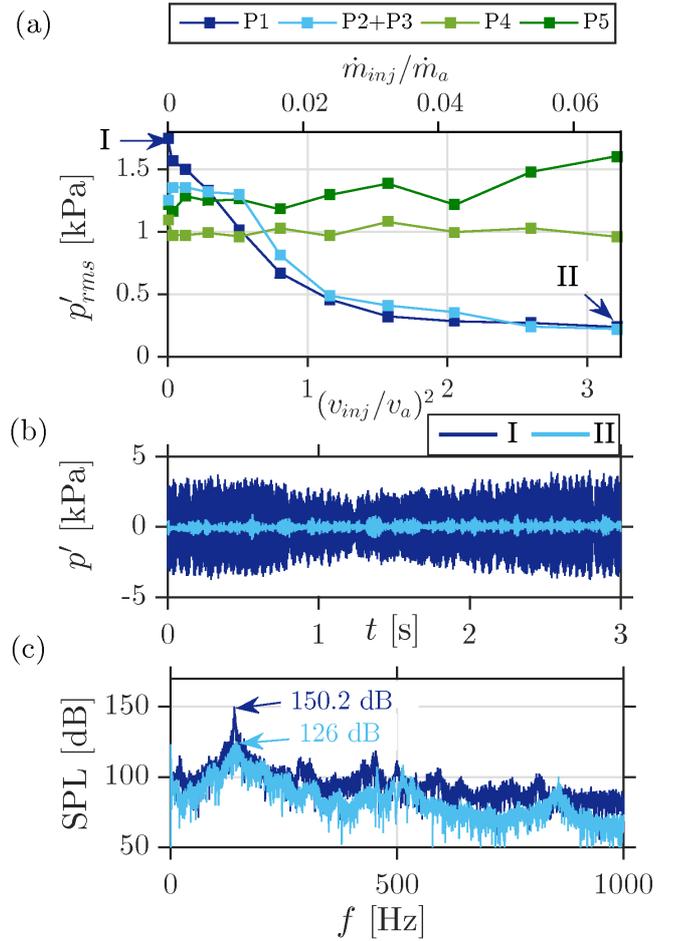}
\caption{\label{Fig6_suppression_plotl} Illustration of control of thermoacoustic instability through secondary air injection targeting different regions of the combustor. (a) Amplitude of pressure fluctuations, $p^\prime_{rms}$, {\color{black}as functions} of the momentum ratio, $(\upsilon_{inj}/\upsilon_a)^2$, and mass flow ratio, $\dot{m}_{inj}/\dot{m}_a$. Representative (b) time series and (c) sound pressure level (SPL) observed during thermoacoustic instability, point I in (a), and subsequent control, point II in (a). Subfigure (a) adapted from \cite{krishnan2019emergence} with permission.}
\end{figure}

Figure \ref{Fig6_suppression_plotl}a shows the amplitude of pressure oscillations ($p^\prime_{rms}$) as a function of the ratio of momentum of the injected air to that of the primary air, $(\upsilon_{inj}/\upsilon_a)^2$ (bottom axis) and the ratio of mass flow rate, $\dot{m}_{inj}/\dot{m}_a$ (top axis). The plot essentially depicts the effectiveness of targeting various regions of the flow-field in getting suppression. We notice that targeting the region between the dump plane and bluff-body through either port P1 or combined injection through P2 \& P3 leads to a significant decrease in $p^\prime_{rms}$. There is a decrease from 1.75 kPa to 0.22 kPa, which amounts to {\color{black}87.4\% reduction in the amplitude of limit cycle oscillations. During the state of control, the amplitude levels are similar to those observed during the state of combustion noise and low-amplitude intermittency}. Individually injecting from P2 and P3 does not lead to any reduction. Further, targeting the region above the bluff-body through port P4 or the region after the bluff-body through port P5 does not lead to any reduction in $p^\prime_{rms}$, and in fact, lead to an increase in the amplitude of limit cycle oscillations. 

We also plot the time series and the sound pressure level (SPL) in Figs. \ref{Fig6_suppression_plotl}b \& \ref{Fig6_suppression_plotl}c at representative points {\color{black}I and II} indicated in Fig. \ref{Fig6_suppression_plotl}a . We observe a shift in the dynamics from the thermoacoustic limit cycle at point {\color{black}I to fluctuations resembling the state of intermittency during suppression at point II. The maximum amplitude during suppression is of the order of 100 Pa.} We further observe a 24.2 dB drop in the sound pressure level from 150.2 dB during thermoacoustic instability to 126 dB after suppression (Fig. \ref{Fig6_suppression_plotl}c). 

The above exercise shows us the relative importance of different regions of the flow field as far as passive control is concerned. The region between the dump plane and bluff-body, as identified by the amplitude of turbulent velocity fluctuations {\color{black}$|\hat{u}_T(f_a)|$} in Fig. \ref{Fig4_CN_INT_TAI}i, is ``critical" to the spatiotemporal dynamics of thermoacoustic instability. Thus, targeting the critical region through P1 and P2+P3 leads to effective control. Other regions such as the top of the bluff-body or region downstream of the bluff-body identified respectively from the Rayleigh index (Fig. \ref{Fig4_CN_INT_TAI}l) and averaged HRR (Fig. \ref{Fig4_CN_INT_TAI}k) are not as crucial and, hence, cannot be used for optimizing the location of secondary injection. This is in direct contrast to Tachibana et al. \cite{tachibana2007active}, who observed the most significant control when secondary micro-jet of air targeted the region of the largest Rayleigh index in their swirl-stabilized combustor.  We surmise that the Rayleigh index identifies the region of the most significant acoustic driving; however, it does not always identify the region most sensitive to control. Thus, we cannot always use the local Rayleigh index or average HRR for determining the critical region. 

\begin{figure}[t!]
\includegraphics[width=\textwidth]{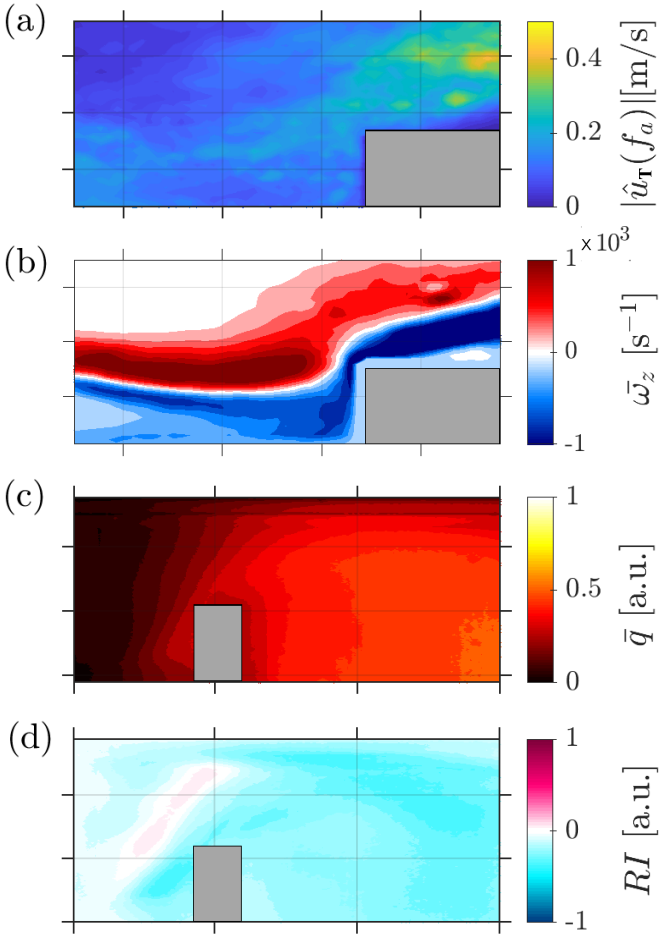}
\caption{\label{Fig7_AfterSuppression} (a) Amplitude of velocity fluctuations {\color{black}$|\hat{u}_T(f_a)|$}, (b) time-averaged vorticity {\color{black}$\bar{\omega}_z$}, (c) time-averaged HRR $\bar{q}$ and (d) Rayligh index $\overline{RI}$ during suppression corresponding to point B in Fig. \ref{Fig6_suppression_plotl}a.}
\end{figure}

\begin{figure*}[t!]
\caption{\label{Fig8_Phase_avg_HRR} Heat release rate field during the state of suppression of thermoacoustic instability for point B indicated in Fig. \ref{Fig6_suppression_plotl}a. (a) Intermittent acoustic pressure fluctuations during suppression and (b) an enlarged portion showing alternate cycles of periodic and aperiodic fluctuations. (c) Points i-iv correspond to $\dot{q}^\prime(x,y)$ at four points of the periodic cycle indicated in (b). Points v-viii {\color{black}correspond} to $\dot{q}^\prime(x,y)$ at the indicated points during aperiodic oscillations as indicated in (b).}
\includegraphics[width=\textwidth]{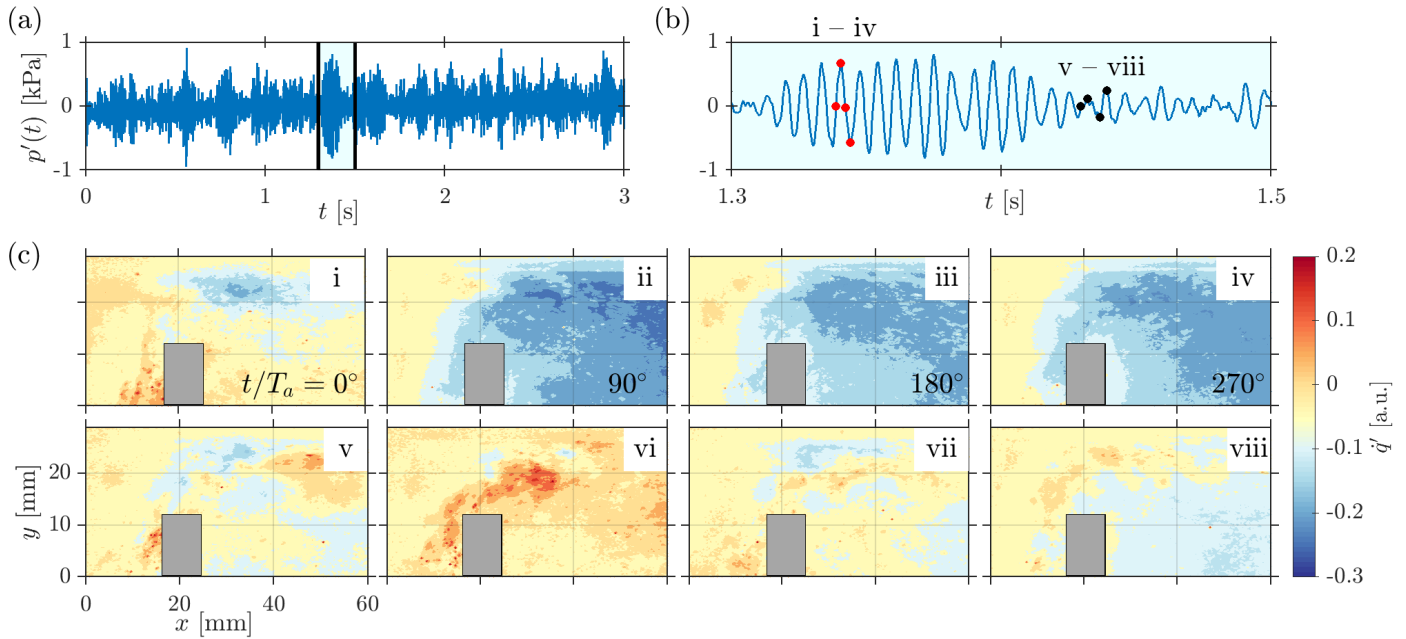}
\end{figure*}

Next, we analyze the effect of secondary air injection on different measures. Figure \ref{Fig7_AfterSuppression} depicts the spatial dynamics associated with the state of suppression (point II indicated in Fig. \ref{Fig6_suppression_plotl}a). We notice many differences from the dynamics during thermoacoustic instability. The maxima of {\color{black}$|\hat{u}_T(f_a)|$} is much lower during the state of control and moves from the region between the dump plane and the bluff-body (Fig. \ref{Fig4_CN_INT_TAI}i) to the top of the bluff-body (Fig. \ref{Fig7_AfterSuppression}a). The time-averaged vorticity field $\bar{\omega}_z$ shows vortices concentrated along a small region along the shear layer, indicating the suppression of the large coherent structures formed during thermoacoustic instability (Fig. \ref{Fig4_CN_INT_TAI}a). We further notice that the mean flame structure (Fig. \ref{Fig7_AfterSuppression}c) is no longer as concentrated downstream of the bluff-body, as was the case during thermoacoustic instability (Figs. \ref{Fig4_CN_INT_TAI}k). In fact, the HRR gets more distributed and extends far downstream. Accordingly, the Rayleigh index is very low value throughout the combustor, indicating the reduction in the strength of acoustic power sources inside the combustor due to secondary injection.

During the state of suppression, the pressure fluctuations show intermittent characteristics (Fig. \ref{Fig8_Phase_avg_HRR}a,b). We plot the instantaneous mean-subtracted HRR field $\dot{q}^\prime(x,y)$ at four points during the epoch of periodic oscillations in Figs. \ref{Fig8_Phase_avg_HRR}ci-iv and aperiodic oscillations in Fig. \ref{Fig8_Phase_avg_HRR}cv-viii. We can observe that the flame is anchored along the shear layer after the bluff-body and extends downstream.  During periodic oscillations, the instantaneous HRR field shows no concentrated spots in the HRR field as were present during thermoacoustic instability (Fig. \ref{Fig5_Q_Ph_Avg.eps}). The entire field is distributed with moderate levels of HRR fluctuations. During aperiodic oscillations, the HRR field is incoherent. Finally, the instantaneous HRR field does not show any visible correlation with pressure fluctuations either during periodic or aperiodic oscillations, leading to very low values of the Rayleigh index (Fig. \ref{Fig7_AfterSuppression}d).  

{\color{black} The suppression of thermoacoustic instability depends upon the underlying mechanism and the effect of micro-jet injection on it. In the present combustor, thermoacoustic instability develops when the reactive field mutually synchronizes with the acoustic field \cite{pawar2017thermoacoustic}. Vortices from the unstable shear layer are shed at the acoustic frequency and develop into large coherent structures in the region between the dump plane and the bluff-body \cite{george2018pattern, premchand2019lagrangian_a}. These large-scale vortices carry unburnt reactants, which upon collision with the bluff-body and the combustor walls, release large amounts of heat (Fig. \ref{Fig5_Q_Ph_Avg.eps}). Consequently, the heat release rate and Rayleigh index are, thus, largest above and beyond the dump plane (Fig. \ref{Fig4_CN_INT_TAI}k,l). However, the source of such a large heat release rate can still be traced back to the unstable shear layer developing at the dump plane \cite{premchand2019lagrangian_b}. 

Secondary injection from port P1 suppresses perturbations in the shear layer from amplifying, thereby hindering the formation of large coherent structures and, eventually, thermoacoustic instability. On the other hand, injection ports P2 and P3 are further downstream of the dump plane and at a larger distance from the shear layer than P1. Thus, injection from P2 and P3 alone are not enough to control the shear layer and require combined injection for control. On the other hand, injection ports P4 and P5 are much further downstream and cannot impede the unstable shear layer from developing into large coherent structures. Thus, they are ineffective in attaining control.}
  
\subsection{Optimized passive control using Hurst exponent}
\label{3.4 Spatial Hurst exponent}
We have established that effective passive control of thermoacoustic instability depends crucially on the region targeted using secondary air-injection. Determination of the critical region is non-trivial as different physical measures point to different regions. Of these, we observed that amplitude of turbulent velocity fluctuations {\color{black}$|\hat{u}_T(x,y,f_a)|$} correctly identifies the critical region, targeting which led to the suppression of thermoacoustic instability. The critical region is also identified from network-based measures, as shown in Krishnan et al. \cite{abin2019critical}. However, neither $|\hat{u}_T(x,y,f_a)|$ nor the network-based measures in \cite{abin2019critical} are able to distinguish the critical region during the state of intermittency. We remedy this by implementing the spatiotemporal analysis using Hurst exponent calculated from the turbulent velocity fluctuations ($u_T$), as discussed in \S\ref{2.3. Tools for analysis}.

\begin{figure}[t!]
\centering
\includegraphics[width=\textwidth]{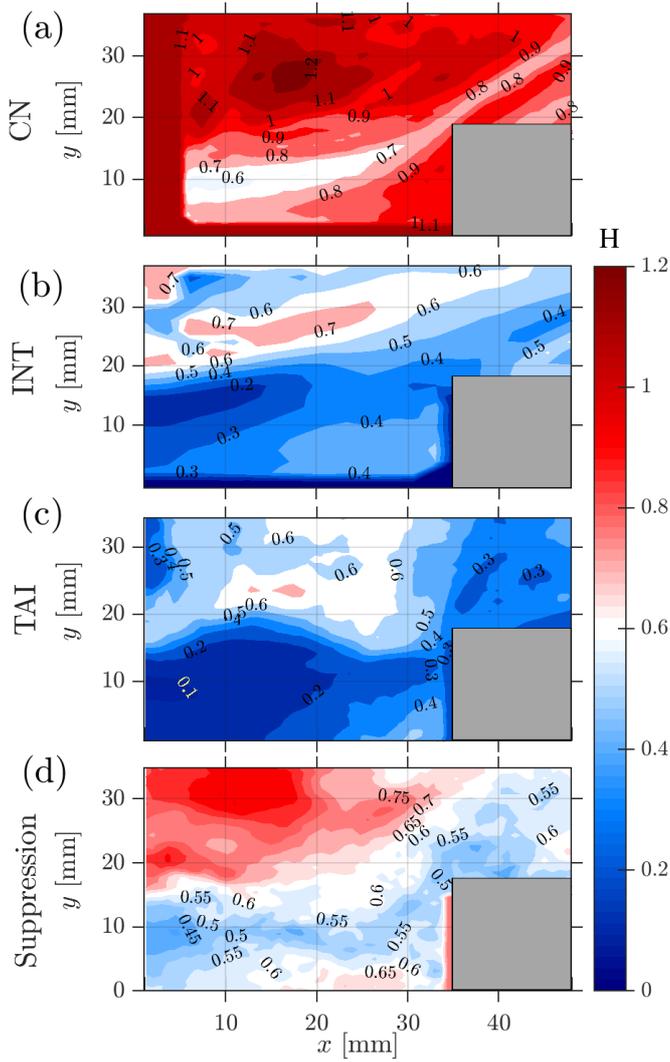}
\caption{\label{Fig9_H_CN_INT_TAI} Field of Hurst exponent ($H$) calculated from the intensity of velocity fluctuations during (a) combustion noise, (b) intermittency and (c) thermoacoustic instability. The experimental conditions for (a-c) correspond to the points A, B and C in Fig. \ref{Fig2_prms_vs_ma}, and (d) corresponds to point II in Fig. \ref{Fig6_suppression_plotl}{\color{black}.}}
\end{figure}

During the state of combustion noise  at $\phi=0.86$ \& $\upsilon_a=8.1$ m/s, $H$ values in the range $[0.5,1.2]$ are distributed throughout the combustor (Fig. \ref{Fig9_H_CN_INT_TAI}a). The regions with different $H$ indicate the difference in the nature of the turbulent velocity fluctuations. The presence of values of $H>0.5$ indicates that the dynamics of velocity fluctuations are persistent, i.e., a large (small) value is more likely to be followed by another large (small) value. In fact, the velocity fluctuations in regions with $0.5<H<1$ have fractal characteristics and possess long-range correlations typically associated with fully-developed turbulent flows \cite{mordant2002long}. Thus, the aperiodically shed small-scale vortices during combustion noise have persistent and fractal behavior.

Figure \ref{Fig9_H_CN_INT_TAI}b shows the distribution of $H$ during the state of intermittency  at $\phi=0.66$ \& $\upsilon_a=11.1$ m/s. We observe large-amplitude periodic bursts embedded randomly amidst the low amplitude aperiodic fluctuations in the measured pressure signal. We notice that the $ H $ values in the range of $[0.2,0.7]$ are distributed across the flow field. Regions with $H<0.5$ indicate anti-persistent behavior in that large (small) values associated with velocity fluctuations are more likely to be followed by a small (large) fluctuation. For $H$ close to 0 indicates anti-persistent and periodic behavior. This indicates that the velocity field exhibits periodicity temporally. The region with $H$ close to 0.2 indicates the region over which spatial coherence is maximum in the field.

\begin{figure}[t!]
\centering
\includegraphics[width=\textwidth]{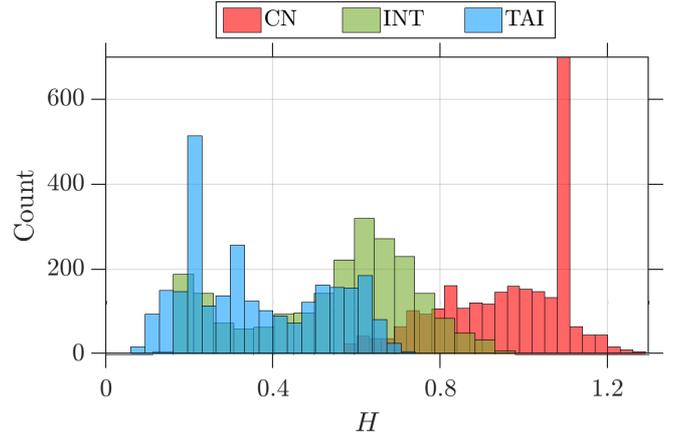}
\caption{\label{Fig10} Histogram showing the distribution of the Hurst exponent (H) during the states of combustion noise (CN), intermittency (INT), and thermoacoustic instability (TAI).}
\end{figure}

During thermoacoustic instability at $\phi=0.63$ \& $\upsilon_a=12.3$ m/s (Fig. \ref{Fig9_H_CN_INT_TAI}c), we observe that the field has $H$ values predominantly {\color{black}lesser} than 0.5. We can observe that the $H$ values in the region immediately after the dump plane is very low ($H\approx0.1$). Signals with $H$ values near-zero imply the absence of scaling of the structure-function. In other words, the fluctuations are bounded. As a consequence, the signal displays a strong anti-persistent behavior, characteristic of periodic signals. Thus, in the region with $H<0.3$, the temporal dynamics of velocity fluctuations are nearly periodic, and much of the spatial region {\color{black}displays} coherence. 

{\color{black} We plot the histogram of $H$ values observed during the three different states in Fig. \ref{Fig10}. We can observe that the histogram shifts to lower $H$ values during the transition from combustion noise to thermoacoustic instability. This shift clearly indicates that the dynamics of velocity fluctuations are predominantly periodic and anti-persistent during thermoacoustic instability. In contrast, during combustion noise, the dynamics show persistent behavior possessing long-range temporal correlations.}

Figure \ref{Fig9_H_CN_INT_TAI}d shows the distribution of $H$ when control using secondary jets from injection port P1 is successfully implemented. The experimental condition corresponds to point II in Fig. \ref{Fig6_suppression_plotl}a. We observe that the expansive region with $H$ close to 0 during the state of thermoacoustic instability, located between the backward-facing step and the bluff-body, is no longer present. In other words, secondary injection disrupted the region with coherence in spatiotemporal dynamics. The resultant flow field has a distribution of $H$ values ranging from 0.5 to 1, which has persistent behavior similar to that observed during the state of combustion noise (Fig. \ref{Fig9_H_CN_INT_TAI}a).
 
Thus, we see that the Hurst exponent correctly identifies the critical region in the flow field, as was also done by Fourier transformed velocity amplitude {\color{black}$|\hat{u}_T(f_a)|$} (Fig. \ref{Fig4_CN_INT_TAI}i). As noted earlier, {\color{black}$|\hat{u}_T(f_a)|$} identified the critical region only during thermoacoustic instability. Similarly, network-based measures also identified the critical region during thermoacoustic instability \cite{abin2019critical}.  In contrast, we observe that $H$ can capture the critical region even during the state of intermittency (Fig. \ref{Fig9_H_CN_INT_TAI}b), which is something we do not observe from the field of {\color{black}$|\hat{u}_T(f_a)|$} during the state of intermittency (Fig. \ref{Fig4_CN_INT_TAI}e). Hence, it is possible to predict the critical region from the velocity field obtained during the state of intermittency. {\color{black}To the best of our knowledge, prediction of the critical region for targeted passive control during the states leading up to thermoacoustic instability has not been achieved until now.} Thus, Hurst exponent proves to be a very robust measure in analyzing the temporal and spatiotemporal dynamics of thermoacoustic systems. 

{\color{black} A literature review suggests that the intermediate state of intermittency is observed in almost all types of combustion systems \cite{nair2014intermittency, gotoda2014detection, pawar2016intermittency, ebi2018flame, kheirkhah2017dynamics, roy2020flame}. Such an intermittent state is also associated with corresponding intermittent states in the spatio-temporal dynamics, such as intermittent vortex shedding \cite{george2018pattern, kheirkhah2017dynamics} and chimera-like states where spatial patches of order develop amidst disorder \cite{mondal2017onset}. Thus, the Hurst exponent can be used to analyze spatial data such as PIV and distinguish spatial regions displaying distinct dynamics. For our purpose, the Hurst exponent was able to predict the critical region where the dynamics are periodic and most sensitive to passive control. We thus believe that the present methodology is generalizable to arbitrary combustor geometries and other intermittent states observed in the combustion dynamics literature.} 

\section{Conclusions}
\label{4. Conclusions}
In the present study, we develop a smart passive control strategy for combating thermoacoustic instability in a bluff-body stabilized turbulent combustor. We analyze the transition from combustion noise to thermoacoustic instability through the state of intermittency. We use several measures, such as the Fourier amplitude of turbulent velocity fluctuations, time-averaged vorticity, averaged heat release rate (HRR), and Rayleigh index to quantify the spatiotemporal dynamics of the combustor. We optimize the secondary air injection location based on the location of the maxima of these physical measures. We find that secondary injection targeting the region of maxima in the amplitude of turbulent velocity fluctuations leads to the highest suppression. We observe a 20 dB drop in the sound pressure level. More importantly, we found that targeting local regions identified from maxima in the Rayleigh index and averaged HRR did not lead to any significant suppression compared to the past study. The Rayleigh index identifies the most significant acoustic power sources; however, it does not identify the most suitable region for control. 

It has been found that the Hurst exponent is a much more robust measure than the velocity amplitude for predicting thermoacoustic instability \cite{nair2014intermittency}. We extend this into the spatiotemporal domain to combine the predictive abilities of Hurst exponent with the idea of optimizing the location of secondary air injection. Using the spatial distribution of the Hurst exponent during the state of thermoacoustic instability, we are able to correctly identify the critical region recognized by the amplitude of turbulent velocity fluctuations, thus validating our approach. We then find that the spatial distribution of Hurst exponent can predict the critical region during the state of intermittency, in contrast to the other physical measures. The capability of the Hurst exponent in predicting the critical region during the state of intermittency constitutes the most important finding of our study. 

In closing, we note that the present methodology has the potential for broad application in combustors. First, it can be used to determine critical regions most suited for aiming control strategies without infringing upon the stability of the flame. Second, the distribution of Hurst exponent can be used to predict critical region if the flow-field is known during the states before full-blown thermoacoustic instability. Thus, for already commissioned combustors, critical regions can be obtained from LES simulations and used to determine the right combination of secondary air-injection ports for efficient control of thermoacoustic instability. Such control can expand the operational regime of combustors leading to reduced costs and energy saving in gas turbine combustors. {\color{black}Finally, we believe that the present methodology can be extended to any turbulent combustion systems displaying intermittent states.}

\section*{Acknowledgments}
This work is funded by Siemens Corporate Technology, India.  We thank Mr. V. Hande, Dr. Renith Richardson, and Dr. Ram Satish Kaluri for their continued support. We thank Mr. Nitin Babu George (PIK, Germany), Ms. K. V. Reeja, Mr. Midhun Raghunathan, Mr. S. Thilagaraj, and Mr. Anand Aravi for their help in conducting the experiments. V. N. and C. P. would like to thank IRCC, IIT Bombay (Grant No. 16IRCCSG006) for funding the study.



\bibliographystyle{elsarticle-num} 
\bibliography{references}

\end{document}